# A gradient index metamaterial

D. R. Smith*,**, J. J. Mock**, A. F. Starr*,***, D. Schurig**

*Department of Mechanical and Aerospace Engineering, University of California, San Diego, La Jolla, CA 92093*

** *Department of Physics, University of California, San Diego, La Jolla, CA 92093*

*** *SensorMetrix, 5965 Pacific Center Blvd., Ste. 701, San Diego, CA 92121-4323*

(Submitted 7 July 2004)

**Abstract**

Metamaterials—artificially structured materials with tailored electromagnetic response—can be designed to have properties difficult to achieve with existing materials. Here we present a structured metamaterial, based on conducting split ring resonators (SRRs), which has an effective index-of-refraction with a constant spatial gradient. We experimentally confirm the gradient by measuring the deflection of a microwave beam by a planar slab of the composite metamaterial over a broad range of frequencies. The gradient index metamaterial represents an alternative approach to the development of gradient index lenses and similar optics that may be advantageous, especially at higher frequencies. In particular, the gradient index material we propose may be suited for terahertz applications, where the magnetic resonant response of SRRs has recently been demonstrated.

There have now been several demonstrations in which electromagnetic material response—either previously unobserved or otherwise difficult to achieve in conventional materials—has been obtained in artificially structured materials [1-4]. These recent demonstrations have shown the potential of artificial materials, often referred to as *metamaterials*, to significantly extend the range of material properties, enabling the potential for new physical and optical behavior, as well as unique electromagnetic devices.

An example of unusual metamaterial response can be found in *negative index metamaterials*, which possess simultaneously negative permittivity ($\varepsilon$) and permittivity ($\mu$) over a finite frequency band. The fundamental nature of negative refraction, hypothesized in 1968 by V. G. Veselago [5], has revealed the key role that metamaterials can play in materials physics, as negative index is a material property not available in existing materials.

The negative index metamaterials thus far demonstrated have been formed from periodic arrays of conducting elements, the size and spacing of which are much less than the wavelengths of interest. The shape of the repeated conducting element determines the electromagnetic response of the collective, which can be approximated as having an electric [6] or a magnetic [7, 8] resonance. Application of effective medium theory to the overall periodically patterned composite allows a description in terms of bulk isotropic or anisotropic $\varepsilon$ and $\mu$.

The *split ring resonator* (SRR), shown in the insets to Fig. 1, has become commonplace as the repeated element in metamaterials that exhibit magnetic properties. A single SRR responds to electromagnetic fields in a manner analogous to a magnetic "atom," exhibiting a resonant magnetic dipolar response. A medium composed of periodically positioned SRRs can be approximately characterized by the following frequency dependent permeability $\mu$:

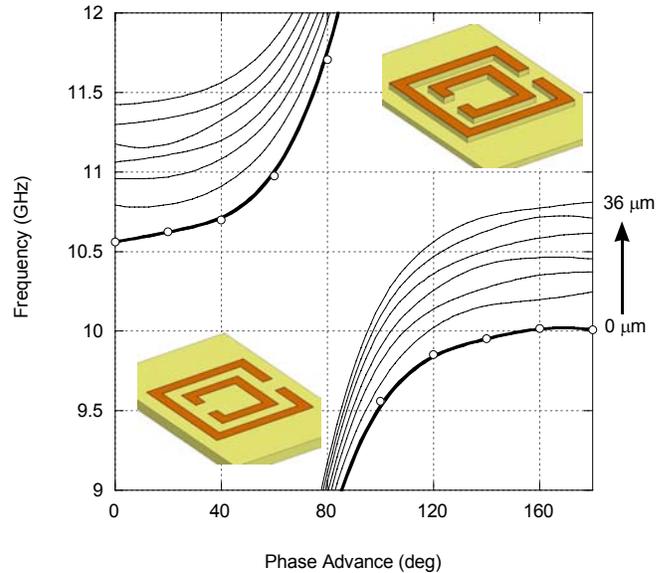

FIG. 1. Simulated dispersion curves for SRRs. The thicker black curve pair (including an upper and lower branch) corresponds to SRRs on a flat substrate (lower inset). The open circles indicate simulated phase advances. Subsequent pairs of curves correspond to cases in which the substrate has been removed around the SRR (upper inset). The depth of cut increases by 6 μm between each set of curves.

$$\mu(\omega) = 1 - \frac{F\omega^2}{(\omega^2 - \omega_r^2) + i\omega\gamma}, \quad (1)$$

where $\omega_r$ is a resonant frequency determined by the SRR geometry, $\gamma$ is the damping and F is the filling factor. The SRR medium also exhibits an effective permittivity, $\varepsilon$, which has been shown to also be dispersive as a function of frequency [9]. However, this frequency dependent behavior is minor at frequencies far away from the resonance, and approaches a constant in the limit of small cell size; thus, we approximate here the permittivity as a constant over frequency. In addition, the orientation of the SRR relative to the field polarization used implies that the electric and magnetic responses are decoupled [10, 11].

In the recent work to date, metamaterials have been constructed from repeated unit cells containing identical elements, such that the resulting medium can be considered homogeneous in the sense that the averaged electromagnetic response does not vary over the structure. However, metamaterials whose averaged electromagnetic properties vary as a function of position can also be fabricated. Such spatially dispersive materials are of interest, for example, as they can be utilized in a variety of applications, including lensing and filtering. Here we present a metamaterial based on SRRs, in which a pattern of spatial dispersion can be introduced by a slight change in the properties of each successive element along a direction perpendicular to the direction of propagation. We thus form a constant gradient index along this axis of the metamaterial, which we confirm by beam deflection experiments.

While an SRR medium is known to have a predominantly magnetic response, this is not of direct interest here; rather, we are concerned with the refractive index n(ω) of the SRR medium, found from $n(\omega) = \sqrt{\varepsilon(\omega)\mu(\omega)}$, with μ(ω) given by Eq. 1 and ε(ω) approximated as constant. This form of the dispersion, ω=ck/n(ω), can be compared with that obtained from a numerical solution of Maxwell's equations for a single unit cell. To obtain the dispersion diagram numerically, we compute the eigenfrequencies for a single unit cell (Fig. 1, inset), applying periodic boundary conditions with zero phase advance in directions perpendicular to the propagation direction, and periodic boundary conditions with various phase advances in the propagation direction. The simulations are performed using HFSS (Ansoft), a finite-element based electromagnetic solver. The resulting dispersion diagram, shown as the frequency versus the phase advance φ across a unit cell (black curve), reveals the expected resonant form. Specifically, there are two branches of propagating modes separated by a frequency band gap. The lower branch starts at zero frequency and ends at $\omega_r$ with a phase advance of 180º. The next branch begins at a frequency $\omega_{mp} = \omega_r / \sqrt{1-F}$ [7]. The propagation constant k can be found from k=φ/d, where d is the size of the unit cell.

The resonant frequency of an SRR, $\omega_r$, depends rather sensitively on the geometrical parameters and local dielectric environment for the SRR [7]. Since μ(ω) depends strongly on $\omega_r$ (Eq. 1), we see that relatively small changes to the basic repeated unit cell can result in substantial changes to the permeability of the composite, especially near the resonance. The change in index $n(\omega) = \sqrt{\varepsilon(\omega)\mu(\omega)}$ with change in resonant frequency can be calculated using Eq. 1. For convenience, we neglect damping; we also set ε(ω)=1, as the primary role of the permittivity over the frequency band of interest will be to rescale the dispersion curves. At low frequencies (ω<<$\omega_r$) the index changes linearly with small changes in the resonance frequency, or

$$\Delta n \sim -\frac{\omega^2}{\omega_r^3}\Delta\omega_r, \qquad (2)$$

whereas in the high frequency limit (ω>>$\omega_r$), we find

$$\Delta n \sim -\frac{\omega_r}{\omega^2}\Delta\omega_r. \qquad (3)$$

Assuming $\Delta\omega_r/\omega_r$<<1 and ignoring higher order terms, we see that, for the model system described by Eq. 1, the gradient increases as the square of the frequency for ω<<$\omega_r$ and decreases as the inverse of the square of the frequency for ω>>$\omega_r$.

There are a variety of modifications to the SRR or its environment that can be used to introduce a variation in $\omega_r$. The method we apply here is to adjust the depth of cut of the dielectric substrate material surrounding the SRR. This method is compatible with our sample fabrication, in which SRRs are patterned on copper clad circuit boards using a numerically controlled micromilling machine. The removal of dielectric material from the region near the SRR (ε~3.8 for FR4 circuit board) changes the local dielectric environment of the SRR, effecting a change in the resonance frequency.

In Fig. 1 we present several dispersion curves corresponding to SRR composites for various depths of substrate material around the SRR. The depth of substrate differs by 6 μm between successive dispersion curves. Fig. 1 shows that $\omega_r$ shifts approximately *linearly* and monotonically with increasing depth of cut, up to 36 μ in depth. Further simulations, as well as the experimental results shown later, show the approximate linearity is valid to 240 μm.

Because the SRR exhibits a resonant frequency $\omega_r$ that increases linearly as a function of the substrate cut depth, it is a convenient element from which to design a gradient index metamaterial. In particular, if we form a metamaterial comprising a linear array of SRRs in which the substrate cut depth advances linearly as a function of cell number, $\omega_r$ will then also advance linearly as a function of cell number; that is, $\omega_r$ becomes linearly proportional to distance. Using this relationship in Eqs. 2 and 3, we see that the gradient of the index will thus be approximately constant as a function of distance, at least for frequencies far enough away from $\omega_r$.

A constant gradient metamaterial can be experimentally confirmed by observing the deflection of a beam incident on a planar metamaterial slab whose index varies linearly (in a direction perpendicular to incident radiation). To calculate this deflection, we consider two normally incident but offset rays entering a gradient index planar slab of thickness t, as shown in Fig. 2. The rays will acquire different phase advances as they propagate through the slab. Assuming the two rays enter at locations x and x+Δx along the slab face, then the acquired phase difference of the two beams traversing the slab,

$$\Phi(x+\Delta x) - \Phi(x) \sim kt \frac{dn}{dx} \Delta x, \quad (4)$$

must equal the phase advance across the path length marked L in Fig. 2. We thus have

$$\sin(\theta) \sim t \frac{dn}{dx} = t \frac{dn}{d\omega_r} \frac{d\omega_r}{d\delta} \frac{d\delta}{dx}, \quad (5)$$

which shows that for a material with a constant spatial gradient in index, the beam is uniformly deflected. Here, $\delta(x)$ is the depth of cut as a function of distance along the slab. This simplified analysis applies strictly to thin samples, as the phase fronts will otherwise not be uniform within the material. Note that $\Phi(x)$ is the phase shift across a slab of arbitrary thickness. If the slab is one unit cell in thickness, then for the SRR cell the phase shift will be $\phi$ as defined earlier.

In order to construct a constant gradient index metamaterial, the above analysis suggests that we form a linear array of SRRs, in which the substrate depth is a linearly increasing function of cell number in the direction perpendicular to the propagation. The resulting array should then deflect an incident beam by an angle that can be predicted by the dispersion diagrams in Fig. 1. To estimate this angle of deflection, we can take the difference between any two of the curves in Fig. 1 to find the gradient of the phase shift per unit cell. The phase shift per unit cell is equivalent to the beam deflection that will be produced by a gradient index metamaterial slab one unit cell thick in the propagation direction. The resulting plot of deflection angle as a function of frequency, obtained from the dispersion curves in Fig. 1, is shown in Fig. 3.

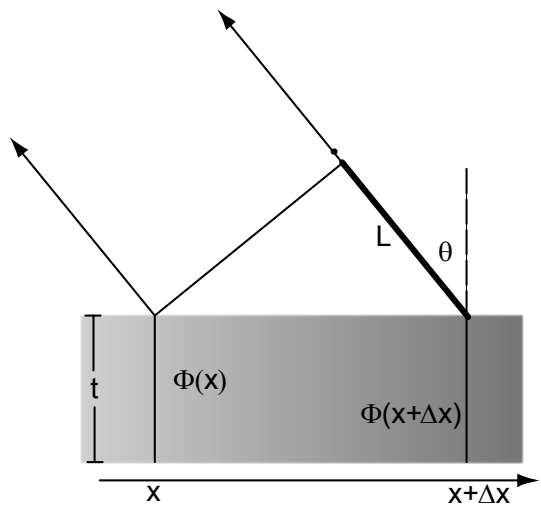

Fig. 2. Diagram showing the deflection of a wave by a structure whose refractive index possesses a gradient that is constant.

The curves in Fig. 3 are useful for calculating deflection angles only for frequencies where the gradient is constant, which can be determined, for example, by analyzing the differences between several of the dispersion curves of Fig. 1. Furthermore, near the resonant frequency on the low side, the absorption resonance leads to a region of anomalous dispersion where the simulation results (which do not take into account losses) are not valid. An additional complicating factor is that the analyzed structure is periodic, so that higher order bands exist at frequencies greater than $\omega_r$ that are not described by Eq. 1. Nevertheless, Fig. 3 provides an indication that at frequencies above the band gap, per unit cell phase shifts of one degree or more should be obtainable from an SRR slab, one unit cell in thickness, in which each successive cell has an additional 6 μm of substrate dielectric removed relative to the previous cell.

To fabricate the gradient index metamaterial samples, we use an LPKF micromilling machine to mill varying length strips of SRRs (number of cells) from copper clad (single side) FR4 circuit board substrates. The SRR design has identical parameters as that reported in [12]. Several samples are fabricated with thicknesses (in the direction of propagation) of 1, 3, or 5 unit cells. The composite metamaterials are composed of roughly forty strips spaced one unit cell apart, each strip having the substrate milled to a different depth. The resonance frequency of each SRR strip is measured in an angular resolved microwave spectrometer (ARMS) described previously [13]. The measured resonance frequencies of each strip versus the depth of cut are plotted in Fig. 4, where the linearity of the fabrication process is confirmed. The nominal difference in substrate thickness between subsequent milling passes was 6 μm. Note that at two depths there are breaks from the linearity; these deviations from linearity coincide with tool bit changes on the milling machine, indicating some lack of reproducibility in repositioning the mill at the nominal zero cut depth position. The resulting linearity, however, proved to be sufficient for the deflection experiment.

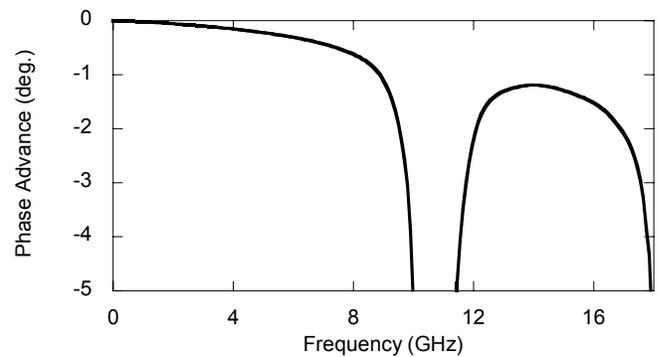

Fig. 3. Frequency versus phase difference per unit cell for the SRR material shown in Fig. 1, in which each successive cell differs by a 6 μm depth of cut.

The composite gradient index samples are measured in the ARMS apparatus. To confirm the gradient in the sample, a microwave beam is directed normally onto the face of the sample (as in Fig. 2), and the power detected as a function of angle at a radius 40 cm away. As in previous measurements on metamaterial samples, the experiment is carried out in a planar waveguide—an effectively two-dimensional geometry in which the electric field is polarized between two conducting (aluminum) plates [13].

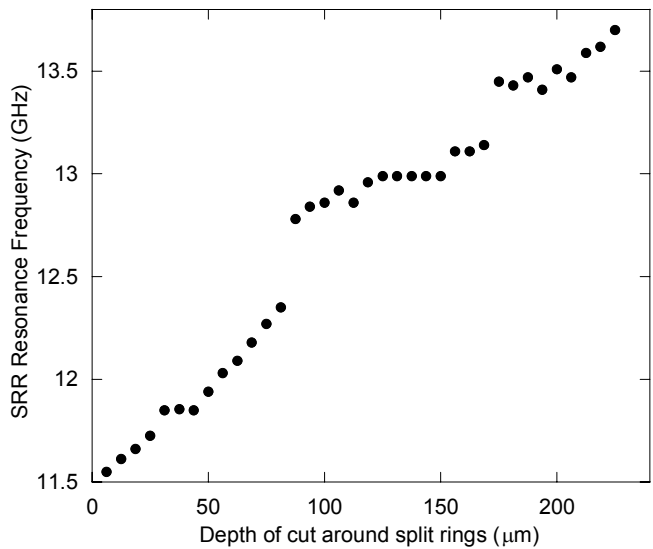

Fig. 4. Resonant frequency versus substrate depth for the machined SRR samples.

In Fig. 5, we present a map of the transmitted power versus angle of detection, as a function of the frequency of the incident microwave beam. Two samples are compared in the figure: Fig. 5a shows a control sample consisting of a five cell deep SRR metamaterial, where each SRR strip is identical (no gradient). The plot in Fig. 5a shows transmission at frequencies corresponding to pass bands, and a frequency region of attenuation corresponding to where the permeability is negative. As can be seen from the figure, the microwave beam exits the sample without deflection, centered about zero degrees.

In Fig. 5b, we present the results of a measurement on an eight cell thick (in propagation direction) gradient index sample, formed by combining the three and five cell samples together. The angular deviation is evident in the figure, especially at the high frequency side of the gap region, where a characteristic tail can be seen in agreement with that predicted in Fig. 3. The qualitative aspects of the curve are in agreement with the theory and simulations above, except that there is weaker evidence of deflection on the low-frequency side of the gap. This lack of symmetry, however, is expected, as the lower frequency side corresponds to the resonance, where the absorption (neglected in the dispersion diagrams) is strongest.

A detailed comparison of the measured and calculated angle of deflection versus frequency is shown in Fig. 6 for the four and eight cell thick gradient index metamaterials.

The curves correspond to the gradient determined from Fig. 3, while the open and black circles are measured points. A frequency translation was applied to the dispersion curve to make the calculated band gap coincident with the band gap measured in the actual structure; no other fitting or adjustments were performed. The excellent agreement shown in Fig. 6 attests to the precision of the fabrication process, as illustrated in Fig. 4. The agreement also provides important evidence that even a single unit cell can be described as having a well-defined refractive index, since the interpretation of this effect depends on a refractive index that varies controllably from cell-to-cell within the structure.

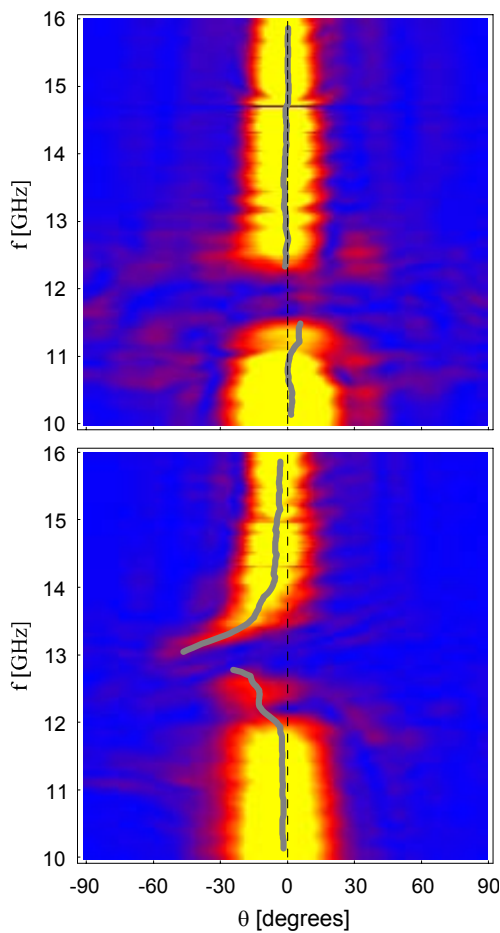

Fig. 5. Angular maps versus frequency of the power emerging from (a) an SRR sample, five unit cells in thickness, with no index gradient and (b) an SRR sample, eight unit cells in thickness, with linear gradient in the direction transverse to the incoming beam.

Figs. 5 and 6 show the practicality of designed spatially dispersive structures. In this case, a linear gradient has been introduced that has the effect of uniformly deflecting a beam by an angle adjustable by design. We find it convenient to work with the SRR system, as the properties of SRRs are now well established. In particular, the resonance frequency of the SRR is relatively easy to identify, can be easily tuned by slightly modifying the parameters of the unit cell, and can be used to roughly parameterize the entire frequency dependence of the SRR. While not the only method for introducing a gradient, the gradient index SRR structure shows the feasibility of creating yet another unique type of metamaterial by combining macroscopic elements.

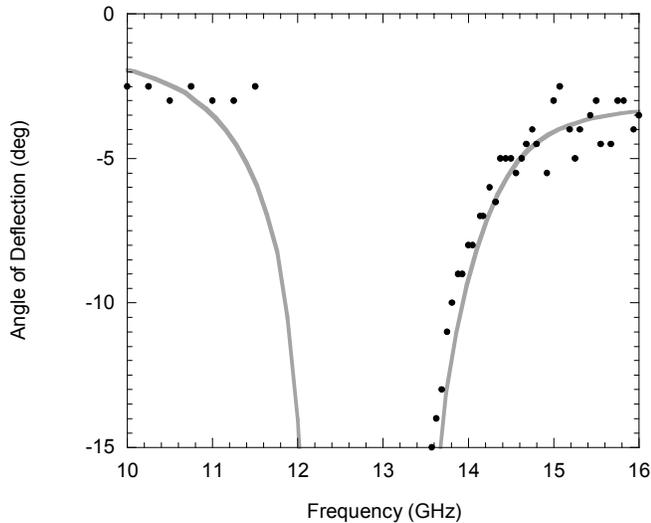

Fig. 6. Measured angle of deflection (black circles) of a gradient index SRR slab, eight unit cells in thickness. The gray curves are taken from those shown in Fig. 3, but have been translated in frequency so that the calculated and the measured band gap regions overlap.

An evident application of the gradient index metamaterial is in gradient index lenses. A parabolic (as opposed to linear) distribution of index in the slab along an axis perpendicular to the direction of wave propagation results in a structure that focuses radiation rather than deflecting it [14]. Common examples of such gradient index lenses include radial gradient index rod lenses, used at optical frequencies, and Luneberg lenses, used at microwave frequencies. Gradient index rod lenses use optical glass materials that are ion doped via thermal diffusion. This process can produce only modest variations of the refractive index, (less than 0.2), and is limited to fairly small diameter rods, (less than 1 cm) [15]. Luneberg spherical or hemispherical lenses, which require the fairly wide index range of n=1 to n=2, can be implemented as stepped index devices with no particular size limitation [14]. Both devices employ gradients in the dielectric permittivity only, and thus have limited impedance matching to the surrounding media.

Gradient index metamaterials may provide a useful alternative approach to the development of optics. The advantages of both traditional and planar lenses formed from artificially patterned media, for example, have been demonstrated at radio and microwave frequencies [16]. With the increased range of material response now identified in metamaterials, including negative refractive index, considerably more flexibility and improved performance from such devices should be possible [17]. Gradient index metamaterials that include magnetic permeability gradients, for example, could be used to develop materials whose index varies spatially but which remains matched to free space. Moreover, with the recent demonstration of SRRs at THz frequencies [18], the gradient index metamaterial should be realizable at higher frequencies.

**Acknowledgements**
This work was supported by DARPA, through an ONR MURI (grant no. N00014-01-1-0803). AFS and DRS also acknowledge support from DARPA through ARO (grant no. DAAD19-00-1-0525).

**References**

1. D. R. Smith, W. Padilla, D. C. Vier, S. C. Nemat-Nasser, S. Schultz, *Phys. Rev. Lett.* **84**, 4184 (2000).
2. R. A. Shelby, D. R. Smith, S. Schultz, *Science*, **292**, 79 (2001).
3. C. G. Parazzoli, R. B. Greegor, K. Li, B. E. C. Koltenbah, M. Tanielian, *Phys. Rev. Lett.*, **90**, 107401 (2003).
4. A. A. Houck, J. B. Brock, I. L. Chuang, *Phys. Rev. Lett.*, **90**, 137401 (2003).
5. V. G. Veselago, *Soviet Physics USPEKHI* **10**, 509 (1968).
6. J. B. Pendry, A. J. Holden, W. J. Stewart, I. Youngs, *Phys. Rev. Lett.* **76**, 4773 (1996).
7. J. B. Pendry, A. J. Holden, D. J. Robbins, W. J. Stewart, *IEEE Trans. MTT*, **47**, 2075 (1999).
8. R. W. Ziolkowski and F. Auzanneau, *J. Appl. Phys.*, **82**, 3192 (1997).
9. T. Koschny, P. Markos, D. R. Smith, C. M. Soukoulis, *Phys. Rev. E*, **68**, 065602 (2003).
10. R. Marques, F. Medina, R. Rafii-El-Idrissi, *Phys. Rev. B*, **65**, 144440 (2002).
11. N. Katsarakis, T. Koschny, M. Kafesaki, E. N. Economou, C. M. Soukoulis, *Appl. Phys. Lett.*, **84**, 2943 (2004).
12. R. A. Shelby, D. R. Smith, S. C. Nemat-Nasser, S. Schultz, *Appl. Phys. Lett.*, **78**, 4 (2001).
13. A. F. Starr, P. M. Rye, J. J. Mock, D. R. Smith, *Rev. Sci. Inst.*, **75**, 820 (2004).
14. E.W. Marchland, *Gradient Index Optics*, Academic Press, New York, (1978).
15. H. Hashizume, K. Hamanaka, A.C. Graham, X.F. Zhu, *Proc. of SPIE*, **4437**, 26, (2001).
16. J. D. Krauss, *Antennas*, 2[nd] Ed. (McGraw-Hill, Boston, 1988), Chapter 14.
17. D. Schurig and D. R. Smith, e-print physics/0403147 (2003).
18. T. J. Yen, W. J. Padilla, N. Fang, D. C. Vier, D. R. Smith, J. B. Pendry, D. N. Basov, X. Zhang, *Science*, **303**, 1494 (2004).